\documentclass[12pt]{article}

\pdfoutput=1

\usepackage{ifpdf}
\ifpdf
\usepackage{graphicx}
\usepackage{hyperref}
\usepackage[dvipsnames]{xcolor}
\else
\usepackage[dvipdfmx]{graphicx}
\usepackage[dvipdfmx]{hyperref}
\usepackage[dvipsnames]{xcolor}
\fi
\hypersetup{
  colorlinks,
  citecolor=magenta,
  linkcolor=Maroon,
  menucolor=black,
  urlcolor=Violet}
\usepackage{amssymb,amsfonts,amsmath,cancel,cite,multirow}
\usepackage{tikz}
\usepackage{zref-clever}
\zcsetup{abbrev = true, cap = true}
\usepackage[margin=1in]{geometry}
\usepackage{bbm}
\usepackage{braket}
\begin{document}

\begin{titlepage}

\begin{flushright}
    % Report Number
\end{flushright}

\vskip 1.35cm
\begin{center}

{\large
\textbf{
    Composite Asymmetric Dark Matter Assisted by Primordial Black Holes Evaporation
}}
\vskip 1.2cm

\mbox{
	Takumi Kuwahara$^{a,b}$ 
	and Yoshiki Uchida$^{c,d,e}$
}

\vskip 0.4cm

\textit{$^a$
    Center for Theoretical Physics and College of Physics, Jilin University, Changchun, 130012, China
}

\textit{$^b$
    Center for High Energy Physics, Peking University, Beijing 100871, China
}

\textit{$^c$
    Institute of Particle Physics and Key Laboratory of Quark and Lepton Physics (MOE), Central China Normal University, Wuhan, Hubei 430079, China
}

\textit{$^d$
    State Key Laboratory of Nuclear Physics and Technology, Institute of Quantum Matter, South China Normal University, Guangzhou 510006, China
}

\textit{$^e$
    Guangdong Basic Research Center of Excellence for Structure and Fundamental Interactions of Matter, Guangdong Provincial Key Laboratory of Nuclear Science, Guangzhou 510006, China
}

\vskip 1.5cm
\begin{abstract}
	\noindent 
    We investigate a cogenesis scenario for composite asymmetric dark matter framework: a dark sector has a similar strong dynamics to quantum chromodynamics in the standard model, and the dark-sector counterpart of baryons is the dark matter candidate.
    The Hawking evaporation of primordial black holes plays the role of a source of heavy scalar particles whose $CP$-violating decay into quarks and dark quarks provides particle--anti-particle asymmetries in baryons and dark matter, respectively. 
    Primordial black holes should evaporate after the electroweak phase transition and before the big-bang nucleosynthesis for explaining the baryon asymmetry of the Universe and for consistent cosmology. 
    We find that this scenario explains the observed values for both baryon and dark matter energy densities when the heavy scalar particles have a mass of $10^6 \text{--} 10^9\, \mathrm{GeV}$ and the primordial black holes have masses of $10^7 \text{--} 10^9\,\mathrm{g}$.
\end{abstract}
\end{center}
\end{titlepage}

%%%%%%%%%%%%%%%%%%%%%%%%%%%%%%%%%%%%%%%%%%%%%%%%%%%%%%%%%%%%%%%%
\section{Introduction}

% Current status of DM and ADM
The existence of dark matter (DM) is firmly confirmed by astrophysical and cosmological observations. 
However, the particle nature of DM remains one of the open questions despite extensive experimental and observational efforts to detect DM signals. 
The production mechanism of DM in the early Universe is one of the key properties determining the DM abundance in the Universe. 
Asymmetric dark matter (ADM) provides an alternative to the standard thermal freeze-out framework for DM, where the particle--anti-particle asymmetry accounts for the DM relic abundance in the current Universe (see Refs.~\cite{Nussinov:1985xr, Barr:1990ca, Barr:1991qn, Kaplan:1991ah, Dodelson:1991iv, Kuzmin:1996he, Fujii:2002aj, Foot:2003jt,Foot:2004pq, Kitano:2004sv, Farrar:2005zd, Gudnason:2006ug, Kitano:2008tk, Kaplan:2009ag} for early work and also Refs.~\cite{Davoudiasl:2012uw, Petraki:2013wwa, Zurek:2013wia} for reviews). 
If the DM number asymmetry is related to the baryon asymmetry of the Universe, the ADM mass can be predicted from the cosmological observations of the visible and DM energy densities. 

% Compositeness
Compositeness plays a significant role in the ADM framework~\cite{Gudnason:2006yj, Dietrich:2006cm, Khlopov:2007ic, Khlopov:2008ty, Foadi:2008qv, Mardon:2009gw, Kribs:2009fy, Barbieri:2010mn, Blennow:2010qp, Lewis:2011zb, Appelquist:2013ms, Hietanen:2013fya, Cline:2013zca, Appelquist:2014jch, Hietanen:2014xca, Krnjaic:2014xza, Detmold:2014qqa, Detmold:2014kba, Asano:2014wra, Brod:2014loa, Antipin:2014qva, Hardy:2014mqa, Appelquist:2015yfa, Appelquist:2015zfa, Antipin:2015xia, Hardy:2015boa, Co:2016akw, Dienes:2016vei, Ishida:2016fbp, Lonsdale:2017mzg, Berryman:2017twh,  Gresham:2017zqi, Gresham:2017cvl, Mitridate:2017oky, Gresham:2018anj, Ibe:2018juk, Braaten:2018xuw, Francis:2018xjd, Bai:2018dxf, Redi:2018muu, Chu:2018faw, Mahbubani:2019pij,Hall:2019rld, Tsai:2020vpi, Asadi:2021yml, Asadi:2021pwo, Zhang:2021orr, Bottaro:2021aal, Ibe:2021gil, Hall:2021zsk, Asadi:2022vkc,Cline:2022leq,Chung:2024nnj,Chung:2024ezq,Chung:2025wle}.
We introduce confining gauge dynamics into the dark sector, where the low-energy theory is described by the hadronic spectrum of the confining dynamics, referred to as dark hadrons. 
A dynamical mass scale for the dark-sector hadrons naturally emerges through the dimensional transmutation, and hence it is possible to explain why the mass scale of dark baryons is close to that of nucleons. 
The dark baryons are stable due to the dark-baryon number conservation analogous to the stability of the proton, and thus the lightest dark baryons are the DM candidate in this framework.
The symmetric component of the DM abundance is strongly depleted due to a sizable annihilation cross section into dark-sector pions (dark pions) in analogy with the strong annihilation of nucleons into pions.
It is necessary to introduce a low-energy portal interaction connecting the two sectors; otherwise, the dark pions may carry most of the energy density after the annihilation. 

% Mechanism for number asymmetries
The mechanisms that relate the two number asymmetries, for baryons and DM, are commonly classified into two types: one involves sharing asymmetries and the other is cogenesis mechanism. 
In the former case, the asymmetry is first generated either in the SM sector or in the dark sector, after which the generated asymmetry is transferred to the other sector (see e.g., Refs.~\cite{Foot:2003jt,Foot:2004pq,Farrar:2005zd,Kaplan:2009ag,Barr:2011cz,Cui:2011qe} for early work).
In order to transfer the generated asymmetry, it is necessary to introduce operators that are singlets under gauge symmetries of both sectors and that carry the baryon (or $B-L$) number and the dark matter number.
In particular, for composite ADM, the portal interaction between the two sectors generally arises from non-renormalizable operators composed of dark quarks~\cite{Ibe:2018juk}, and hence ultraviolet (UV) completions have also been explored~\cite{Ibe:2018tex,Ibe:2019ena,Lonsdale:2018xwd,Murgui:2021eqf,Chung:2024nnj}.
Meanwhile, a single source can generate both the baryon asymmetry and the DM asymmetry in the cogenesis framework (see e.g., Refs.~\cite{Kitano:2004sv,Cosme:2005sb,Kitano:2008tk,Davoudiasl:2010am,Gu:2010ft,Falkowski:2011xh} for early work). 
In the literature, it is often assumed that a single particle couples to both baryon (or lepton) and DM, and its decay generates both asymmetries. 
One may consider a macroscopic object as a source of asymmetries, such as the decay of a scalar configuration into both baryon and DM~\cite{Cheung:2011if,vonHarling:2012yn} (for example, known as Affleck-Dine mechanism~\cite{Affleck:1984fy}) and evaporation of the primordial black holes (PBHs)~\cite{Barman:2021ost}.

% Cogenesis scenario for composite ADM
The PBHs are hypothetical compact objects formed at the early stage of the Universe~\cite{Carr:1974nx,Carr:1975qj}, due to the cosmological energy contrast triggered by inflation (see e.g., Refs.~\cite{Carr:1993aq,Yokoyama:1998pt,Garcia-Bellido:2017mdw,Ballesteros:2017fsr,Hertzberg:2017dkh,Martin:2019nuw,Martin:2020fgl} for single field inflation), phase transition (see e.g., Refs.~\cite{Hawking:1982ga,Kodama:1982sf,Sasaki:1982fi,Jedamzik:1999am,Lewicki:2019gmv,Gouttenoire:2023naa,Lewicki:2023ioy,Hashino:2025fse,Wang:2025hwc}), extended objects, and so on (see Ref.~\cite{Carr:2016drx} for instance).
Evaporation of light PBHs may trigger the generation of the baryon asymmetry of the Universe~\cite{Barrow:1990he,Baumann:2007yr,Fujita:2014hha,Hook:2014mla,Hamada:2016jnq,Hooper:2020otu,Datta:2020bht}.
In this study, we propose a cogenesis scenario for the composite ADM framework, where the PBHs in a specific mass range trigger the generation of both baryon and dark-matter asymmetries.
It is natural that PBHs serve as a common origin of the particles responsible for generating both asymmetries, because gravity couples universally and does not discriminate between the standard model (SM) sector and the dark sector. 
As the first step of this scenario, the PBHs can produce very heavy particles at the last stage of the PBH evaporation via the Hawking radiation~\cite{Hawking:1974rv,Hawking:1975vcx}, even if the temperature of the Universe is lower than the masses of these particles.
The subsequent asymmetry generation resembles the conventional grand unification (GUT) baryogenesis~\cite{Dimopoulos:1978kv,Yoshimura:1978ex,Weinberg:1979bt,Toussaint:1978br} and the conventional leptogenesis~\cite{Fukugita:1986hr}.
The out-of-equilibrium decay of the heavy particles, with baryon (or lepton) number violation and $C$ and $CP$ violation, into light particles generates the particle--anti-particle asymmetries.
Indeed, the PBHs with the mass range of $10^6\text{--}10^9\,\mathrm{g}$ complete their evaporation after the decoupling of the sphaleron process but before the big-bang nucleosynthesis (BBN).
Therefore, it is possible to explain the baryon asymmetry of the Universe directly from the decay of heavy particles (as in GUT baryogenesis) without conflicting with other cosmological constraints~\cite{Hooper:2020otu}.
Given that the dark sector mirrors the SM structure in the composite ADM framework, PBHs can also provide the heavy particles responsible for generating the DM particle--anti-particle asymmetry. 
This cogenesis scenario reproduces the observed energy densities when the heavy particles have masses of $10^6 \text{--} 10^9\, \mathrm{GeV}$.

The outline of this paper is as follows.
We briefly discuss the PBHs and summarize our notation in \zcref{sec:PBHs}, while we discuss our particle-physics model, composite ADM, in \zcref{sec:cADM}.
We discuss a baryogenesis model via the PBH evaporation in \zcref{sec:PBHsAsym}, and discuss the prediction of the ADM mass in this framework.
\zcref{sec:Conc} is devoted to concluding this study.

%%%%%%%%%%%%%%%%%%%%%%%%%%%%%%%%%%%%%%%%%%%%%%%%%%%%%%%%%%%%%%%%
\section{Primordial Black Holes and Evaporation \label{sec:PBHs}}

We begin by briefly describing PBH properties and PBH evaporation, establishing our notation used throughout this study. 
The formation and mass of PBHs are model dependent; there are several ways to form PBHs, such as primordial density perturbation from inflation, phase transition, and so on.
We discuss the PBH properties under simple assumptions, as the analysis is not expected to be significantly affected by the precise details of the PBH formation.
We assume that the PBHs do not have charge or angular momentum. 
Although we consider a dark-sector scenario in this study, we assume that the dark sector remains in thermal equilibrium with the SM through low-energy portal interactions.

Once spherical PBHs form during the radiation-dominated era, the initial mass of the PBH is close to the horizon mass enclosed within the Hubble volume at the formation time, which is given by 
\begin{align}
    M_\mathrm{BH}^\mathrm{in} 
    = \gamma \, \rho_R \, \frac{4\pi}{3H^3} \,, 
\end{align}
where $\rho_R \equiv 3 H^2 M_\mathrm{Pl}^2$ denotes the energy density of the radiation, $H$ is the Hubble rate, $M_\mathrm{Pl} = 2.4 \times 10^{18} \, \mathrm{GeV}$ is the reduced Planck mass, and $\gamma \simeq 0.2$ is a numerical factor depending on the details of the gravitational collapse~\cite{Carr:2020gox}. 
Although the constraints on PBHs change for the extended mass function~\cite{Carr:2017jsz}, we assume a monochromatic mass function for the PBH mass for simplicity throughout this study.
We define the ratio of the initial energy densities of radiation and PBHs as follows.
\begin{align}
    \beta \equiv \frac{\rho_\mathrm{BH}(T_0)}{\rho_R(T_0)}\,, 
\end{align}
where the initial energy density of PBHs is given by $\rho_\mathrm{BH}(T_0) = n_\mathrm{BH}^\mathrm{in} M_\mathrm{BH}^\mathrm{in}$ with the initial number density of PBHs, $n_\mathrm{BH}^\mathrm{in}$. 
Since we assume that the PBHs form in the radiation-dominated era, we obtain the temperature of the Universe at the PBH formation, denoted by $T_0$, as  
\begin{align}
    T_0 = 2 \left(\frac{10}{g_{\ast\rho}}\right)^{1/4} \sqrt{\frac{3 \gamma M_\mathrm{Pl}^3}{M_\mathrm{BH}^\mathrm{in}}}
    \simeq 1.3 \times 10^{12} \, \mathrm{GeV} \left(\frac{106.75}{g_{\ast\rho}}\right)^{1/4} \left(\frac{10^7 \, \mathrm{g}}{M_\mathrm{BH}^\mathrm{in}}\right)^{1/2} \,.
    \label{eq:T0}
\end{align}
Here, $g_{\ast\rho}$ denotes the effective number of degrees of freedom, contributing to the radiation component of the energy density, at the PBH formation.
Since the ratio of the energy densities of radiation and PBHs scales with the scale factor (i.e., $\rho_\mathrm{BH}/\rho_R \propto a$), the PBHs may dominate the energy density of the Universe before evaporation completes.
Indeed, once we assume that the effective degrees of freedom do not drastically change during the cosmic evolution, the energy densities of radiation and PBHs approximately evolve as $\rho_R(T)/\rho_R(T_0) \simeq T^4/T_0^4$ and $\rho_\mathrm{BH}(T)/\rho_\mathrm{BH}(T_0) \simeq T^3/T_0^3$, respectively. 
One can estimate the critical value for the initial abundance ratio $\beta$ such that the PBHs dominate the energy density of the Universe before the completion of their evaporation. 
If we ignore the temperature dependence of $g_{\ast \rho}$, the PBH lifetime $\tau_\mathrm{PBH}$ and the reheating temperature $T_\mathrm{RH}$ due to evaporation are approximately given by \cite{Papanikolaou:2020qtd}
\begin{align}
    \tau_\mathrm{PBH} \simeq \frac{160}{\pi \, g_{\ast \rho}} \frac{(M_\mathrm{BH}^{\mathrm{in}})^3}{M_\mathrm{Pl}^4} \,, \quad 
    T_\mathrm{RH} \simeq \left( \frac{90 \, g_{\ast \rho}}{10240} \right)^{1/4} \frac{M_\mathrm{Pl}^{5/2}}{(M_\mathrm{BH}^\mathrm{in})^{3/2}}  \,.
\end{align}
The ratio of the energy densities, $\rho_\mathrm{BH}(T)/\rho_R(T)$, reaches its maximum just before the completion of the evaporation due to the scale factor dependence. 
Therefore, if the ratio at the reheating temperature $T_\mathrm{RH}$ exceeds order unity, $\rho_\mathrm{BH}(T_\mathrm{RH})/\rho_R(T_\mathrm{RH}) \gtrsim 1$, PBHs can dominate the energy density of the Universe when the PBH evaporation ends. 
This condition gives the minimum value of the initial energy-density ratio required for the presence of the matter-dominated epoch during the cosmic history, 
\begin{align}
    \beta \simeq \frac{T_\mathrm{RH}}{T_0} \frac{\rho_\mathrm{BH}(T_\mathrm{RH})}{\rho_R(T_\mathrm{RH})} \gtrsim \beta_\mathrm{min}  \simeq 3 \times 10^{-12} \left( \frac{10^6 \, \mathrm{g}}{M_\mathrm{BH}^\mathrm{in}} \right) \left( \frac{g_{\ast \rho}}{106.75} \right)^{1/2} \,.
\end{align}
Here, we use the parameter dependence of the temperature at the PBH formation, $T_0$, given in \zcref{eq:T0}, and we define $\beta_\mathrm{min} \equiv T_\mathrm{RH}/T_0$.
Even if the initial abundance of PBHs is very small compared to that of radiation, PBHs readily come to dominate the energy density of the Universe before evaporating. 
We note that PBHs may produce stochastic gravitational waves even if they evaporate before the BBN epoch~\cite{Anantua:2008am,Zagorac:2019ekv,Inomata:2020lmk,Papanikolaou:2020qtd}.
Our scenario can remain viable even if the initial abundance and mass lie well outside the sensitivity of future observations.
As stated, we do not specify the formation mechanism of PBHs in the early Universe in this study, and hence we take the ratio $\beta$ to be a free parameter that is below the future sensitivities to the gravitational wave observations. 
However, it is important to take these sensitivities into account once we specify the mechanism for the PBH formation.
 
Now, we discuss the particle production from the PBH evaporation in detail. 
A particle is emitted by PBHs as Hawking radiation when its mass is below the black hole (BH) temperature, 
\begin{align}
    T_\mathrm{BH} 
    = \frac{M_\mathrm{Pl}^2}{M_\mathrm{BH}} \simeq 1.05 \times 10^7 \, \mathrm{GeV} \left( \frac{10^6 \, \mathrm{g}}{M_\mathrm{BH}} \right) \,.
\end{align}
A heavy particle is produced at the last stage of the evaporation since $T_\mathrm{BH}$ increases as the PBH loses mass. 
The PBH loses mass through particle emission, and the mass-loss rate of an evaporating PBH is expressed as~\cite{Carr:2009jm,Carr:2020gox}
\begin{align}
    \frac{d M_\mathrm{BH}}{dt} \equiv - 5.34 \times 10^{25} \epsilon(M_\mathrm{BH}) \left(\frac{1 \, \mathrm{g}}{M_\mathrm{BH}} \right)^2 \mathrm{g \, sec^{-1}} \,.
    \label{eq:massloss}
\end{align}
Here, a function $\epsilon(M_\mathrm{BH})$ measures the number of emitted particles \cite{MacGibbon:1991tj}.
The SM contribution for the function $\epsilon(M_\mathrm{BH})$ is evaluated as \cite{MacGibbon:1990zk,MacGibbon:1991tj}
\begin{align}
    \epsilon_\mathrm{SM}(M_\mathrm{BH})
    & = 2 f_1 + 6 f_{1/2}^0 
    + 4 f_{1/2}^1 \left( \sum_\ell e^{-\frac{M_\mathrm{BH}}{\beta_{1/2} M_\ell}} + 3 \sum_q e^{-\frac{M_\mathrm{BH}}{\beta_{1/2} M_q}}\right) \nonumber \\
    & \quad + 16  f_1 e^{-\frac{M_\mathrm{BH}}{\beta_1 M_g}} 
    + 3 f_1 \left( 2 e^{-\frac{M_\mathrm{BH}}{\beta_1 M_W}} + e^{-\frac{M_\mathrm{BH}}{\beta_1 M_Z}}\right) + f_0 e^{-\frac{M_\mathrm{BH}}{\beta_{0} M_H}} \,.
    \label{eq:eps_SM}
\end{align}
We normalize the function $\epsilon(M_\mathrm{BH})$ to unity, $\epsilon_\mathrm{SM}(M_\mathrm{BH}) = 1$, as the BH mass $M_\mathrm{BH} \gg 10^{17} \, \mathrm{g}$, corresponding to the emission of only (almost) massless particles, photon and three generations of neutrinos.
Each term in \zcref{eq:eps_SM} describes the contributions from photon, neutrinos, charged leptons and quarks, gluons, electroweak gauge bosons, and the physical Higgs boson from the first term. 
The coefficient for each contribution represents the inner degrees of freedom, such as particle-antiparticle, spin, and colors. 
The constants $\beta_i$ and $f_i^{(q)}$ with spin $i$ are 
\begin{align}
    \beta_0 = 2.66 \,, \quad
    \beta_{1/2} = 4.53 \,, \quad
    \beta_{1} = 6.04 \,.
\end{align}
and 
\begin{align}
    f_s =
    \begin{cases}
        0.267 & \quad (s = 0) \\
        0.060 & \quad (s = 1) 
    \end{cases} \,, \quad 
    f_{1/2}^{q} =
    \begin{cases}
        0.147 & \quad (q = 0:\text{neutral}) \\
        0.142 & \quad (q = 1:\text{charged}) 
    \end{cases} \,.
\end{align}
The mass parameter $M_i$ for particle $i$ denotes the BH mass whose temperature corresponds to the particle mass $m_i$, 
\begin{align}
    M_i = \frac{M_\mathrm{Pl}^2}{m_i} \simeq 10^{13} \mathrm{g} \left(\frac{1 \mathrm{GeV}}{m_i}\right) \,.
    \label{eq:evap_mass}
\end{align}
As for light quarks and gluons, we choose the effective masses to be their dynamical masses $m_q = 300 \, \mathrm{MeV}$ and $m_g = 600 \, \mathrm{MeV}$, respectively, as taken in Refs.~\cite{MacGibbon:1990zk,MacGibbon:1991tj}.
The production of massive particles is exponentially suppressed until the BH temperature gets larger than their masses, as shown in \zcref{eq:eps_SM}.
In practice, we stop solving \zcref{eq:massloss} once the PBH mass reaches the Planck mass $4 \pi M_\mathrm{Pl}$ (see \zcref{{app:anal_PBH}} for detail). 
There may exist stable remnants of the PBH evaporation, known as the Planck-mass relics~\cite{MacGibbon:1987my,Barrow:1992hq,Carr:1994ar,Chen:2004ft}, which can contribute to the DM abundance. 
We ignore the presence of the Planck-mass relics in this study, since we remain agnostic about the impact of the quantum gravity on the final stage of evaporation.

We also ignore a possible back-reaction from the emitted particle on the Hawking evaporation.
A back-reaction from the emitted particles on the quantum state of BHs would become important when the energy of the emitted particles is close to the BH's total energy. 
This effect, known as memory-burden effect~\cite{Dvali:2018xpy,Dvali:2024hsb}, may slow down the evaporation by resisting the loss of the PBH mass. 
If present, it could modify the allowed region for the PBH mass due to the altered evaporation history. 
It would be worthwhile to investigate the impact of the memory-burden effect on the baryon and DM asymmetries, which we leave for future work.

Next, we discuss how PBHs affect the evolution of the energy density of the Universe. 
We define the comoving energy densities of the PBHs and the radiation as $\varrho_\mathrm{BH} \equiv a^3 \rho_\mathrm{BH}$ and $\varrho_R \equiv a^4 \rho_R$ with the scale factor $a$, respectively.
The energy densities evolve following the equations~\cite{Perez-Gonzalez:2020vnz,JyotiDas:2021shi,Bernal:2022pue}, 
\begin{align}
    a H \frac{d \varrho_R }{da} & = - f_R \frac{d \ln M_\mathrm{BH}}{dt} a \varrho_\mathrm{BH} \,, \\ 
    a H \frac{d \varrho_\mathrm{BH} }{da} & = \frac{d \ln M_\mathrm{BH}}{dt} \varrho_\mathrm{BH} \,, \label{eq:diffeq-varrhoPBH}\\
    H^2 & = \frac{1}{3 M_\mathrm{Pl}^2} \left(\frac{\varrho_\mathrm{BH}}{a^3} + \frac{\varrho_R}{a^4} \right) \,.
\end{align}
Here, $f_R$ denotes the fraction of Hawking radiation composed of light particles, which contribute to the radiation component of the energy density of the Universe, $f_R = \epsilon_R(M_\mathrm{BH})/\epsilon(M_\mathrm{BH})$. 
The left-hand sides of these equations are proportional to the mass-loss rate of the PBH given by \zcref{eq:massloss}.
In particular, the comoving energy density for radiation remains constant in the absence of PBHs (i.e., $\varrho_\mathrm{BH} = 0$); namely, the radiation energy scales $\rho_R \propto a^{-4}$. 
Meanwhile, since the PBH evaporation after the matter-dominated era reheats the Universe, the comoving entropy density, $\mathcal{S} \equiv s a^3$, also increases by following the equation,
\begin{align}
    a H \frac{d \mathcal{S}}{da} & = - f_R \frac{d \ln M_\mathrm{BH}}{dt} \frac{\varrho_\mathrm{BH}}{T} \,.
    \label{eq:entropy_injection}
\end{align}
The comoving entropy density is conserved as $\varrho_\mathrm{BH} \to 0$ as expected. 
In this study, we include the impact of the entropy injection, which reduces the final yields of heavy particles just after the PBH evaporation (this effect can also be crucial for the high-scale leptogenesis~\cite{Calabrese:2023key}).
In addition, PBH evaporation modifies the scale-factor dependence of the temperature of the thermal plasma as follows. 
\begin{align}
    \frac{a H}{T}\frac{dT}{da} 
    & = - \frac{1}{\Delta} \left( H + \frac{f_R}{4} \frac{d \ln M_\mathrm{BH}}{dt} \frac{g_{\ast\rho}(T)}{g_{\ast S}(T)}\frac{\rho_\mathrm{BH}}{\rho_R} \right) \,, \quad 
    \Delta \equiv 1 + \frac{T}{3 g_{\ast S}} \frac{d g_{\ast S}}{dT} \,.
    \label{eq:temp_evol}
\end{align}
Here, $g_{\ast S}$ counts the effective number of the degrees of freedom contributing to the radiation component of the entropy density. 
We take into account the temperature dependence of the effective degrees of freedom contributing to the energy density $g_{\ast \rho}$ and to the entropy density $g_{\ast S}$, in particular we include the SM contribution taken from Ref.~\cite{Saikawa:2020swg}.

After we introduce the composite ADM framework in the next section, we discuss the cogenesis scenario utilizing the PBH evaporation in \zcref{sec:PBHsAsym}.

%%%%%%%%%%%%%%%%%%%%%%%%%%%%%%%%%%%%%%%%%%%%%%%%%%%%%%%%%%%%%%%%
\section{Composite Asymmetric Dark Matter \label{sec:cADM}}

In this section, we briefly review the composite-particle realization of ADM framework.
We establish the notation and discuss the particle contents of the model. 

We consider a dark-sector model with confining $SU(N_c)_D$ gauge dynamics, which is commonly referred to as dark quantum chromodynamics (QCD). 
In this study, we take $N_c = 3$ for simplicity.%
\footnote{
    Since the color number of two sectors is the same in this case, we can consider the mirror model as a UV model for the dark sector (see e.g., \cite{Ibe:2018tex,Ibe:2019ena,Lonsdale:2018xwd,Murgui:2021eqf,Chung:2024nnj} for mirror unification).
    This setup explains why the dynamical scale of the dark QCD $\Lambda_\mathrm{QCD'}$ is close to that of the QCD: the strong couplings in both sectors are related to each other at the UV scale, and similar renormalization evolution of the strong couplings leads to similar dynamical scales for dark QCD and SM QCD. 
}
We introduce $N_f$-flavor vector-like quarks (referred to as dark quarks) in the (anti)fundamental representation of $SU(3)_D$. 
$N_f$ denotes the number of flavors active at low energy, whereas additional dark quarks may decouple from the low-energy spectrum due to sizable vector-like masses. 
There is an upper bound on the number of flavors required to preserve the asymptotic freedom of the dark QCD, and hence we assume a moderate number of flavors $N_f$ in this study. 
The vector-like masses for the $N_f$-flavor dark quarks are assumed to be universal, collectively denoted by $m_{q'}$, and below a dynamical scale of the dark QCD, denoted by $\Lambda_\mathrm{QCD'}$.
The low-energy theory of dark QCD below the dynamical scale is described by composite particles (dark hadrons), and there are two kinds of dark hadrons: dark mesons and dark baryons. 
Since we have an approximate chiral symmetry for dark quarks, the lightest dark mesons are dark pions, pseudo Nambu-Goldstone bosons of the chiral symmetry breaking in the dark sector, and the masses of such dark pions are of order $(m_{q'} \Lambda_\mathrm{QCD'})^{1/2}$.
Meanwhile, the mass of dark baryons is dynamically generated and of order $\Lambda_\mathrm{QCD'}$. 
If the dark QCD coupling is comparable to the SM QCD coupling at some high energy scale, we expect that the dark dynamical scale $\Lambda_\mathrm{QCD'}$ to be close to the QCD scale $\Lambda_\mathrm{QCD} \simeq 200 \,\mathrm{MeV}$.

\begin{table}
    \caption{ 
        Charge assignment of dark-sector particles in a composite ADM model: Weyl dark quarks and dark Higgs. 
        $SU(3)_D$ and $U(1)_D$ are gauge symmetries of the dark sector.
        \label{tab:part_cont}
    }
    \centering
    \begin{tabular}{|c|c|c|}
        \hline
        & $SU(3)_D$ & $U(1)_D$ \\ \hline
        $u'$ & $\mathbf{3}$ & $2/3$ \\
        $\overline u'$ & $\overline{\mathbf{3}}$ & $-2/3$ \\
        $d'$ & $\mathbf{3}$ & $-1/3$ \\
        $\overline d'$ & $\overline{\mathbf{3}}$ & $1/3$ \\ \hline
        $H'$ & $\mathbf{1}$ & $1$ \\ \hline
    \end{tabular}
\end{table}

By analogy with the stability of protons, the lightest dark baryon is stable due to the accidental conservation of the dark baryon number, making it a viable DM candidate.
However, by analogy with the SM, most of dark baryons generated in the early Universe do not survive in the current Universe due to strong annihilation into dark pion. 
Therefore, a particle--anti-particle asymmetry in dark baryons must be generated to account for the DM abundance.
If the dark baryon and baryon asymmetries are related, the DM mass can be predicted from the observation of the energy densities of baryons and DMs. 
\begin{align}
    \frac{\rho_\mathrm{DM}}{\rho_B} = \frac{m_\mathrm{DM} \eta_\mathrm{DM}}{m_B \eta_B} \simeq 5.45 \,.
\end{align}
Here, $\eta_{B\,, \mathrm{DM}}$ denotes the number asymmetry of baryons and DM. 
In other words, in the composite ADM framework, the dark-sector dynamical scale $\Lambda_\mathrm{QCD'}$ must be close to that of the SM sector if $\eta_{B} \simeq \eta_{\mathrm{DM}}$.

Furthermore, the strong annihilation of dark baryons produces a substantial abundance of dark pions, and hence the dark pions would also behave as DM if stable and would carry a large entropy density in the Universe and contribute as dark radiation~\cite{Blennow:2012de}.
To avoid such cosmological problems, we often introduce a portal interaction connecting two sectors to release significant energy through decay and annihilation of dark pions. 
In this study, we introduce a portal mediated by a massive dark photon that kinematically mixes with the SM photon.
Through the same dark photon, the dark sector can remain in the thermal equilibrium with the SM sector.
The massive dark photon arises as the $U(1)_D$ gauge boson, which acquires its mass through the Higgs mechanism triggered by the dark Higgs $H'$. 
The dark quarks are assumed to have the $U(1)_D$ charges analogous to the electromagnetic charges of the SM quarks. 
The dark pions can be depleted into dark photons through decay and annihilation by the $U(1)_D$ interactions, and then dark photons decay into the SM fermions via the kinetic mixing.
To realize this, we assume the dark photon is the lightest particle in the dark sector. 
The particle contents and the charge assignments are summarized in \zcref{tab:part_cont}.

%%%%%%%%%%%%%%%%%%%%%%%%%%%%%%%%%%%%%%%%%%%%%%%%%%%%%%%%%%%%%%%%
\section{Particle Asymmetries via Primordial Black Holes \label{sec:PBHsAsym}}

Analogous to the baryon asymmetry of the Universe, a particle--anti-particle asymmetry in the dark sector must be generated to account for the DM abundance, since dark baryons are strongly depleted into dark pions. 
In this section, we study the generation of particle--anti-particle asymmetries in both the SM sector and the dark sector via the out-of-equilibrium, $CP$-violating decay of heavy scalar particles produced from the PBH evaporation. 
Even though we consider an analogy of the GUT baryogenesis, the heavy scalar particles whose decay generate the particle asymmetry may not be related to the GUT scale physics. 
Indeed, their masses can be much lower than the GUT scale of $10^{16}\,\mathrm{GeV}$, and we do not further investigate the UV origin of the heavy scalar particles in this study.

We introduce scalar particles $\mathcal{T}$ and $\mathcal{T}'$ in the fundamental representation of $SU(3)_C$ and $SU(3)_D$, respectively.%
\footnote{
    Instead of introducing heavy scalars in each sector, we may consider bi-charged scalar particles that couple to both quarks and dark quarks. 
    We comment on this case in \zcref{sec:Conc}.
}
The scalar particles $\mathcal{T}$ ($\mathcal{T}'$) couple to quarks (dark quarks), and their masses are much larger than the QCD (dark QCD) dynamical scale of about $\mathcal{O}(1)\,\mathrm{GeV}$.
Various assignments of $U(1)$ charges are possible for $\mathcal{T}$ and $\mathcal{T}'$, which allow the Yukawa interactions to (dark) quarks: precisely, the hypercharge of $\mathcal{T}$ and the $U(1)_D$ charge of $\mathcal{T}'$.  
When we choose their charges to be $- 1/3$, the possible Yukawa interactions among the heavy scalar particles and (dark) quarks (in terms of two-component Weyl spinors) are%
\footnote{
    We may consider different $N_c$ from the number of the SM colors.
    We expect that a similar analysis will be applicable even in such a case by introducing a scalar field in the antisymmetric tensor representation allowing us a coupling to two dark quarks, such as $\mathcal{T}'_{[\alpha',\beta']} u^{\prime\alpha'} d^{\prime\beta'}$.
} 
\begin{align}
    \mathcal{L}_\mathrm{int} 
    & = y_{\mathcal{T}_i 1}^L \epsilon_{\alpha \beta \gamma} \epsilon^{rs} \mathcal{T}_i^\alpha Q^{\beta}_{r} Q^{\gamma}_{s} 
    + y_{\mathcal{T}_i 1}^R \mathcal{T}_i^\alpha \overline u_{\alpha} \overline e
    + y_{\mathcal{T}_i 2}^L \mathcal{T}^\ast_{i \alpha} Q^\alpha L
    + y_{\mathcal{T}_i 2}^R \epsilon^{\alpha \beta \gamma} \mathcal{T}^\ast_{i \alpha} \overline u_{\beta} \overline d_{\gamma} 
    \nonumber \\
    & \quad + y_{\mathcal{T}'_{i'} 1} \epsilon_{\alpha' \beta' \gamma'} \mathcal{T}_{i'}'^{\alpha'} u^{\prime\beta'} d^{\prime\gamma'}
    + y_{\mathcal{T}'_{i'} 2} \epsilon^{\alpha' \beta' \gamma'} \mathcal{T}'^\ast_{i' \alpha'} \overline u'_{\beta'} \overline d'_{\gamma'}
    + y_{\mathcal{T}'_{i'} 3} \mathcal{T}'^\ast_{i' \alpha'} d^{\prime\alpha'} \psi
    + \mathrm{h.c.} \,,
    \label{eq:lag_int}
\end{align}
where greek indices (with primes) represent the (dark) color indices,  roman indices $r\,,s$ represent the weak indices, and $i \,, i'$ denote the indices for the scalar particles. 
The flavor indices for both quarks and dark quarks are implicit. 
We also introduce a gauge-singlet fermion $\psi$ in the dark sector, which has the Majorana mass term and is decoupled from the low energy spectrum by its Majorana mass.
The baryon number and the dark baryon number are violated in the presence of the Yukawa interactions, and the $CP$ violation is introduced by the complex phases in the Yukawa couplings.

The Yukawa structure is constrained by the following considerations: 
(i) it must not wash out the generated asymmetries in both the baryon number and the dark baryon number, and 
(ii) it must ensure the stability of the lightest baryon and the lightest dark baryon. 
When the decay of the heavy scalar particles produce the same amount of (dark) baryon and (dark) antibaryon, the net asymmetries are washed out.
The Yukawa couplings violate the (dark) baryon number, it may lead to the instability of proton (lightest dark baryon).
In particular, the Yukawa couplings of $\mathcal{T}$ to the first-generation quarks and leptons are required to be small in order to suppress nucleon decay mediated by $\mathcal{T}$.
The nucleon decay, such as $p \to e^+ \pi^0$, is strongly constrained by the Super-Kamiokande experiment~\cite{Super-Kamiokande:2020wjk}.
The nucleon decay rate via $\mathcal{T}$, which is the lightest $\mathcal{T}_i$, is given by 
\begin{align}
    \Gamma(p \to e^+ \pi^0) 
    & \simeq \frac{1}{64 \pi} (|y_{\mathcal{T} 1}^R y_{\mathcal{T}2}^R|^2 + |y_{\mathcal{T} 1}^L y_{\mathcal{T}2}^L|^2)\frac{m_p}{m_\mathcal{T}^4} |W|^2 \nonumber \\
    & \simeq (4.2 \times 10^{36} \, \mathrm{years})^{-1} \left( \frac{y_\mathcal{T}}{10^{-10}} \right)^4 \left( \frac{10^6 \, \mathrm{GeV}}{m_\mathcal{T}} \right)^4 \,,
\end{align}
where $y_\mathcal{T}$ collectively denotes the size of Yukawa couplings, and $W$ denotes the nuclear matrix elements, $|W| \simeq 0.1 \,\mathrm{GeV}^2$.
It is assumed that the Yukawa coupling of $\mathcal{T}$ is nothing to do with that of the SM Higgs doublet at low energy.
At the UV scale, it may not lead to unification of $\mathcal{T}$ and the SM Higgs doublet into a single representation of the GUT gauge symmetry. 
Once we ignore the unification of $\mathcal{T}$ and the SM Higgs doublet, there are several ways to avoid the nucleon decay constraints.
For example, if $\mathcal{T}$ very feebly couples to leptons (or does not couple to leptons), proton decay is quite suppressed (or proton cannot decay into any fermions) by mediating $\mathcal{T}$. 
The other possibility is to suppress the Yukawa couplings of $\mathcal{T}$ to the first-generation quarks enough for avoiding nucleon decay at low energy: of course, not only the direct couplings to the first-generation quarks but also those to the second and third generations must be small enough to suppress the nucleon decay rate via the Cabibbo-Kobayashi-Maskawa mixing.
It is beyond the scope of this study to discuss the UV origin of the Yukawa structure, and we assume that the Yukawa couplings of $\mathcal{T}$ leading to the nucleon decay are small enough to suppress the nucleon decay rate below the current experimental bound.
Meanwhile, the lightest dark baryon (namely, DM particle) is stable unless we introduce the dark-sector counterpart of leptons. 

It is known that two or more scalar particles are required in the radiative correction contributing to the decay process for generating particle asymmetry (see Refs.~\cite{Botella:1990vf,Liu:1993tg} for GUT baryogenesis and Ref.~\cite{Covi:1996wh} for leptogenesis).
A nonvanishing particle asymmetry can be generated even when the masses of the scalar particles are hierarchical. 
In this study, we assume that $\mathcal{T}$ ($\mathcal{T}'$) is the lightest among color-triplet scalars (dark color-triplet scalars).
$\mathcal{T}$ ($\mathcal{T}'$) is the only dynamical degree of freedom responsible for generating the asymmetry in each sector.
Since we can assign the $B-L$ charge to $\mathcal{T}$ in a consistent manner in \zcref{eq:lag_int}, the $B-L$ is conserved in the interactions involving $\mathcal{T}$.
Therefore, unless the initial asymmetry in the $B-L$ is generated in the early Universe, the most part of the baryon asymmetry generated by the thermal $\mathcal{T}$ decay is washed out by the sphaleron process.
There is no counterpart of the sphaleron process in the dark sector, and hence the dark baryon asymmetry generated by the thermal $\mathcal{T}'$ decay is not washed out.
However, as we saw in \zcref{sec:PBHs}, the thermal contributions to the asymmetries are quite suppressed due to the large energy injection from the PBH evaporation, and hence we do not take into account the asymmetries generated by the thermal decay of $\mathcal{T}$ and $\mathcal{T}'$ in the following.
We illustrate the thermal contributions to the asymmetries in \zcref{app:thermal_contribution}.

The heavy scalar particles $\mathcal{T}$ and $\mathcal{T}'$ are produced as a part of radiation during the last stage of the PBH evaporation, and their energy densities evolve following the equations 
\begin{align}
    a H \frac{d \varrho_{\mathcal{T}^{(\prime)}} }{da} & = - f_{\mathcal{T}^{(\prime)}} \frac{d \ln M_\mathrm{BH}}{dt} a \varrho_\mathrm{BH} \,, 
    \label{eq:rho_T}
\end{align}
with the comoving energy densities for $\mathcal{T}$ and $\mathcal{T}'$, denoted by $\varrho_{\mathcal{T}\,,\mathcal{T}'}$.
The fraction evaporating into $\mathcal{T}$ and $\mathcal{T}'$ is described by $f_{\mathcal{T}^{(\prime)}} = \epsilon_{\mathcal{T}^{(\prime)}}(M_\mathrm{BH})/\epsilon(M_\mathrm{BH})$ where $\epsilon_{\mathcal{T}^{(\prime)}}(M_\mathrm{BH})$ is the contribution from $\mathcal{T}^{(\prime)}$ to the evaporation function $\epsilon(M_\mathrm{BH})$.
The function $\epsilon(M_\mathrm{BH})$ includes extra contributions from the dark-sector particles and the heavy colored scalar particles, $\epsilon(M_\mathrm{BH}) = \epsilon_\mathrm{SM}(M_\mathrm{BH}) + \epsilon_\mathrm{DS}(M_\mathrm{BH}) + \epsilon_\mathrm{h}(M_\mathrm{BH})$. 
The function $\epsilon_\mathrm{SM}$ is given by \zcref{eq:eps_SM}, and the other functions are 
\begin{align}
    \epsilon_\mathrm{DS}(M_\mathrm{BH})
    & = 12 f_{1/2}^1 \sum_{q'} e^{-\frac{M_\mathrm{BH}}{\beta_{1/2} M_{q'}}} 
    + 16 f_1 e^{-\frac{M_\mathrm{BH}}{\beta_1 M_{g'}}} 
    + 3 f_1 e^{-\frac{M_\mathrm{BH}}{\beta_1 M_{A'}}} 
    + f_0 e^{-\frac{M_\mathrm{BH}}{\beta_{0} M_{H'}}} \,, \\
    \epsilon_\mathrm{h}(M_\mathrm{BH})
    & = 6 f_0 e^{-\frac{M_\mathrm{BH}}{\beta_{0} M_\mathcal{T}}}
    + 6 f_0 e^{-\frac{M_\mathrm{BH}}{\beta_{0} M_{\mathcal{T}'}}} \,.
\end{align}
Here, $\epsilon_\mathrm{DS}$ denotes the contribution from light particles in the dark sector, while $\epsilon_\mathrm{h}$ denotes the contribution from the heavy scalar particles, color-triplets $\mathcal{T}$ and dark color-triplet $\mathcal{T}'$.
The mass parameter $M_i$ for the particle $i$ again denotes the BH mass whose temperature corresponds to the particle mass $m_i$, see \zcref{eq:evap_mass}.
In particular, the mass parameters for dark quarks and dark gluon have the relation 
\begin{align}
    M_{q'} = M_q \times \left( \frac{m_B}{m_\mathrm{DM}} \right) \,, \quad 
    M_{g'} = M_g \times \left( \frac{m_B}{m_\mathrm{DM}} \right) \,, 
\end{align}
since the ratio of the dynamical scales in both sectors can be written in terms of the mass ratio of baryon and DM. 
We set the masses of dark photon and dark Higgs to be $m_{A'} \simeq m_{H'} \simeq 100 \, \mathrm{MeV}$.
The DM particles (precisely dark quarks) are produced through evaporation, but no particle--anti-particle asymmetry would not be generated via evaporation.
Hence, the DM abundance from PBH evaporation is sufficiently suppressed due to the strong depletion of dark baryons, and we can ignore the (symmetric) abundance from evaporation.%
\footnote{
    There is another possibility to produce (symmetric component of) DM particles from PBHs~\cite{Gunn:2024xaq}, from the PBH hot spot~\cite{Das:2021wei,He:2022wwy,He:2024wvt}.
    The PBH evaporation can generate very energetic DM~\cite{Fujita:2014hha,Allahverdi:2017sks,Lennon:2017tqq,Morrison:2018xla,Masina:2020xhk} that would affect the structure formation if its population is sufficient enough (and solely gravitational interaction)~\cite{Baldes:2020nuv,Chen:2025gnm}.
}

Since the function $\epsilon_{\mathcal{T}^{(\prime)}}(M_\mathrm{BH})$ is exponentially suppressed until the PBH mass is below the threshold given by \zcref{eq:evap_mass}, the heavy particles, $\mathcal{T}$ and $\mathcal{T}'$, are produced at the last stage of the PBH evaporation. 
The yields of the scalar particles are given by 
\begin{align}
    Y_{\mathcal{T}^{(\prime)}}(T) = \frac{1}{m_{\mathcal{T}^{(\prime)}}} \frac{\rho_{\mathcal{T}^{(\prime)}}(T)}{s(T)} \,. 
\end{align}
Then, the yields for the baryon asymmetry and the DM asymmetry are 
\begin{align}
    Y_B(T) = \varepsilon_\mathcal{T} Y_\mathcal{T}(T) \,, \quad 
    Y_\mathrm{DM}(T) = \varepsilon_{\mathcal{T}'} Y_{\mathcal{T}'}(T) \,.
\end{align}
Here, $\varepsilon_\mathcal{T}$ and $\varepsilon_{\mathcal{T}'}$ denote the $CP$-violating parameters for the decay of $\mathcal{T}$ and $\mathcal{T}'$, respectively. 
As we mentioned, the Yukawa couplings of $\mathcal{T}$ and $\mathcal{T}'$ are assumed to be nothing to do with the Yukawa couplings of the SM Higgs doublet.
The $CP$-violating parameters $\varepsilon_{\mathcal{T}^{(\prime)}}$ are just free parameters of this scenario. 
We find the resulting mass ratio of DM and baryon as follows.
\begin{align}
    \frac{m_\mathrm{DM}}{m_B} & = \frac{\rho_\mathrm{DM}}{\rho_B} \cdot \frac{\eta_B}{\eta_\mathrm{DM}} = 5.45 \frac{\varepsilon_\mathcal{T} \rho_\mathcal{T} m_{\mathcal{T}'}}{\varepsilon_{\mathcal{T}'} \rho_{\mathcal{T}'} m_{\mathcal{T}}} \,.
    \label{eq:DMmass}
\end{align}
Here, we use $\rho_\mathrm{DM}/\rho_B = \Omega_\mathrm{DM} h^2/\Omega_B h^2 \simeq 5.45$ and $\eta_\mathrm{DM}/\eta_B = Y_\mathrm{DM}/Y_B$.

\begin{figure}[!t]
  \centering
    \includegraphics[width=0.75\linewidth]{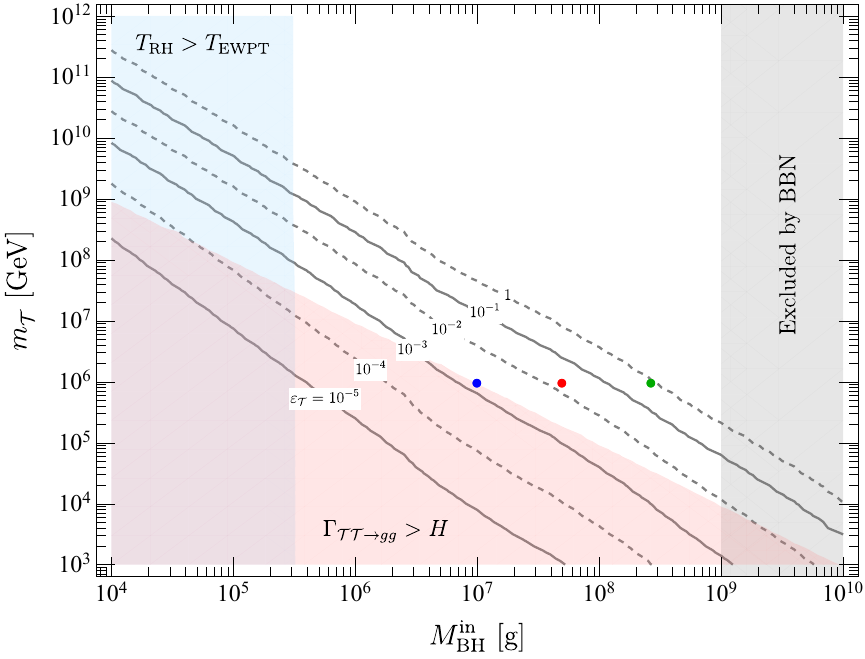}
  \caption{
    The $M_\mathrm{BH}^\mathrm{in}$--$m_\mathcal{T}$ plot for the successful baryogenesis: we obtain the observed baryon energy density $\Omega_B h^2 = 0.022$ on black lines (different values of $\varepsilon_\mathcal{T}$ are written on each lines).
    The reheating temperature above the electroweak scale $T_\mathrm{EWPT} = 160 \, \mathrm{GeV}$ (the light-cyan region) requires the primordial $B-L$ asymmetry. 
    The light-gray region is excluded by the late-time evaporation of PBHs (corresponding mass of $M_\mathrm{BH} \geq 10^9 \, \mathrm{g}$), and the pink shaded region is excluded by the strong depletion of the produced $\mathcal{T}$.
    There are several dots on the plot, which correspond to the reference points used in the following figures. 
}
  \label{fig:mPBHvsmT}
\end{figure}

For now, we examine the asymmetry generation only in the SM sector from the out-of-equilibrium decay of the heavy scalar particle $\mathcal{T}$, assuming the absence of the dark-sector particles. 
We choose the initial abundance of PBHs to be $\beta = 10^2 \beta_\mathrm{min}$ to ensure that PBHs dominate the energy density of the Universe before evaporation. 
The results are insensitive to the choice of $\beta$ as long as PBHs are formed during the radiation-dominated epoch and the Universe enters a matter-dominated epoch before evaporation.
This is because the energy density $\rho_\mathrm{DM}$ follows the same time evolution as the energy density $\rho_R$ in the absence of PBHs just after the energy densities of matter and radiation are equal (see \zcref{app:beta_dependence} for details).
In \zcref{fig:mPBHvsmT}, we show the viable parameter space for achieving the observed baryon energy density $\Omega_B h^2 = 0.022$ for the $CP$-violating parameter $\varepsilon_\mathcal{T}$ ranging from $10^{-5}$ to $1$. 

The Universe is reheated by the emission of relativistic particles from PBHs, and we define the reheating temperature $T_\mathrm{RH}$ as the temperature when the evaporation ends.
When the reheating temperature exceeds the sphaleron decoupling temperature, the sphaleron process re-enters the thermal equilibrium and washes out the generated baryon asymmetry. 
A primordial $B-L$ number asymmetry is therefore required for explaining the observed baryon asymmetry of the Universe, which is depicted as the light-cyan shaded area.
One can consider the non-thermal leptogenesis scenario through the generation of primordial $B-L$ number via the PBH evaporation, and the PBH mass should be below $3 \times 10^5 \, \mathrm{g}$ to convert the primordial $B-L$ number to the baryon asymmetry through the sphaleron process.
Meanwhile, the PBH evaporation continues even during the BBN if the PBH mass exceeds $ 10^9 \, \mathrm{g}$. 
In such a case, the emitted particles from the PBH evaporation affects the primordial abundances of the light nuclei~\cite{Carr:2020gox}, and thus this region is excluded (depicted by the light-gray shaded region).%
\footnote{
    Improvement BBN constraints may extend this exclusion down to the PBH mass of $\mathcal{O}(10^8) \, \mathrm{g}$~\cite{Keith:2020jww,Boccia:2024nly}.
}

When $\mathcal{T}$ is sufficiently light, it is efficiently depleted by the annihilation into gluons. 
The annihilation rate is given by the product of the annihilation cross section and the number density of $\mathcal{T}$, 
\begin{align}
    \Gamma_{\mathcal{TT} \to gg} = \sigma(\mathcal{TT} \to gg) v n_\mathcal{T} \,, \quad 
    \sigma(\mathcal{TT} \to gg) \simeq \frac{7 \pi \alpha_s^2}{m_\mathcal{T}^2} \,.
\end{align}
Here, the number density just after the PBH evaporation is $n_\mathcal{T} = \rho_\mathcal{T}(T_\mathrm{RH})/m_\mathcal{T}$. 
The annihilation process should be decoupled immediately after evaporation; therefore, the annihilation rate should be below the Hubble rate $H$ at the time (namely, $H = 1/2t_\mathrm{RH}$ with the time $t_\mathrm{RH}$ when the evaporation ends).
The region of light $m_\mathcal{T}$ excluded by efficient depletion of $\mathcal{T}$ is depicted as the pink-shaded region in \zcref{fig:mPBHvsmT}.
As a summary, the parameter space compatible with the observed baryon asymmetry is limited, and $\varepsilon_\mathcal{T} \gtrsim 10^{-4}$ is required to generate the observed baryon asymmetry through the PBH evaporation. 

We now compare our treatment of the number density of heavy particles with the previous approaches in the literature. 
We use the cosmological evolutions of the energy density of $\mathcal{T}$ as a part of radiation, given in \zcref{eq:rho_T}, to compute the number density of $\mathcal{T}$ after the PBH evaporation. 
It is apparent that a portion of the energy possessed by the PBHs is transferred to the heavy scalar particles. 
The (comoving) energy density (instead of the number density) is independent of its mass $m_\mathcal{T}$ for small $m_\mathcal{T}$, and hence the resulting baryon asymmetry depends on $m_\mathcal{T}$ even for small $m_\mathcal{T}$. 

Alternatively, one may calculate the total number of the heavy scalar $\mathcal{T}$ from integrating over a statistical (Bose-Einstein) distribution for radiative particles with the Hawking temperature as discussed in Refs.~\cite{Baumann:2007yr,Fujita:2014hha,Hooper:2020otu}.
The total production number is given by 
\begin{align}
    \frac{d N_\mathcal{T}}{d t}
    = \pi r_s^2 \mathcal{G}g_{H}^{\mathcal{T}} 
    \int \frac{d^3 p}{(2\pi)^3} \frac{1}{e^{E/T_\mathrm{BH}}-1} \,,
\end{align} 
where $r_s \equiv 2 M_\mathrm{BH}/M_\mathrm{Pl}^2$ denotes the Schwarzschild radius, $\mathcal{G} \simeq 3.8$ is the appropriate greybody factor, and $g_{H}^{\mathcal{T}}$ denotes the number of the effective degrees of freedom of $\mathcal{T}$.
In this approach, the produced number of $\mathcal{T}$ is independent of its mass $m_\mathcal{T}$ when the initial BH temperature, $M_\mathrm{Pl}^2/M^\mathrm{in}_\mathrm{BH}$, exceeds its mass $m_\mathcal{T}$. 
Hence, the resulting baryon asymmetry no longer depends on $m_\mathcal{T}$ for small $m_\mathcal{T}$, which contrasts to our result.

\begin{figure}[t]
  \centering
    \includegraphics[width=0.45\linewidth]{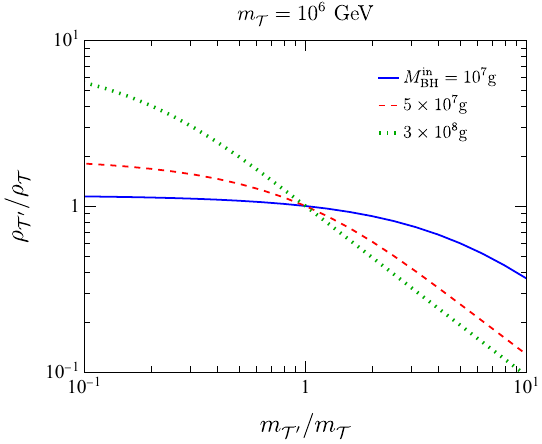}
    \includegraphics[width=0.45\linewidth]{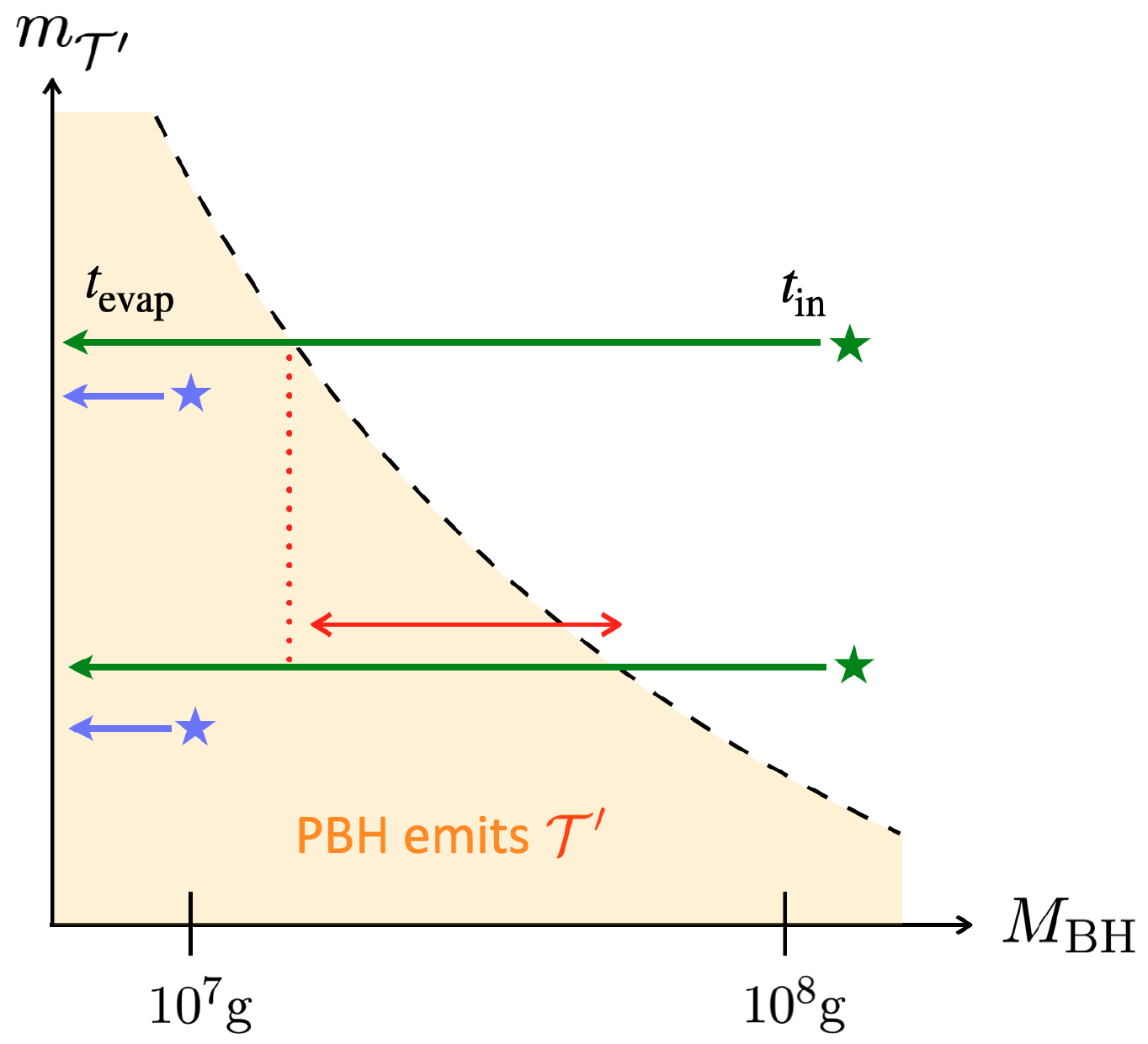}
  \caption{
    (\textit{Left}): The energy density ratio $\rho_{\mathcal{T}'}/\rho_{\mathcal{T}}$ at the reheating temperature as a function of $m_{\mathcal{T}'}/m_{\mathcal{T}}$ with fixed $m_\mathcal{T} = 10^6 \,\mathrm{GeV}$: different input PBH masses $M_\mathrm{BH}^\mathrm{in} = 10^7 \, \mathrm{g}$ (blue thick), $M_\mathrm{BH}^\mathrm{in} = 5 \times 10^7 \, \mathrm{g}$ (red dashed), and $M_\mathrm{BH}^\mathrm{in} = 3 \times 10^8 \, \mathrm{g}$ (green dotted). 
    (\textit{Right}): A schematic picture for the energy density ratios with different PBH mass. 
    The black-dashed line depicts $m_{\mathcal{T}'}$ that is equal to the PBH temperature as a function of the PBH mass $M_\mathrm{BH}$. 
  }
  \label{fig:rhoTratio}
\end{figure}

Next, we introduce the dark-sector particles and examine the DM mass prediction for given energy density ratio $\rho_\mathrm{DM}/\rho_B = \Omega_\mathrm{DM} h^2/\Omega_B h^2 \simeq 5.45$. 
We take the number of dark-quark flavors to be $N_f = 3$ in the numerical analysis.
The results are qualitatively insensitive to the choice of $N_f$ as far as the dark quarks are confined at low energy for the following reasons. 
The energy densities of $\mathcal{T}$ and $\mathcal{T}'$ vary with the evaporation fractions, denoted by $f_{\mathcal{T}^{(\prime)}}$, and the change of $N_f$ affects only a part of total amount of the radiation component of the energy density of the Universe. 
The impact of $N_f$ on the results can be negligible compared to the other parameters such as the heavy scalar masses and the PBH mass. 

The right panel of \zcref{fig:rhoTratio} shows a schematic picture for $m_{\mathcal{T}'}/m_\mathcal{T}$ dependence of the energy density ratio $\rho_{\mathcal{T}'}/\rho_{\mathcal{T}}$. 
In the figure, $m_{\mathcal{T}'}$ is equal to the PBH temperature, $m_{\mathcal{T}'} = T_\mathrm{BH} (M_\mathrm{BH})$, on the black-dashed line, and PBHs can emit $\mathcal{T}'$ in the cream-yellow area.
The star symbols in the figure correspond to the initial mass of PBHs ($3 \times 10^8 \, \mathrm{g}$ for the green points and $10^7 \, \mathrm{g}$ for the blue points), and the PBHs lose their mass via the Hawking radiation along each arrow.
The PBH mass enters the cream-yellow area when it falls down below a certain threshold, resulting the eventual production of $\mathcal{T}'$. 
For a large PBH mass $M_\mathrm{BH}$, the lighter scalar particle starts being produced earlier during the evaporation process than the heavier scalar particle (the delay of the production period is shown as the red arrow). 
As a result, the final energy density of the lighter scalar is larger than that of the heavier scalar.

The left panel of \zcref{fig:rhoTratio} shows the ratio of the energy densities just after the PBH evaporation is completed $\rho_{\mathcal{T}'}(T_\mathrm{RH})/\rho_{\mathcal{T}}(T_\mathrm{RH})$ as a function of $m_{\mathcal{T}'}/m_\mathcal{T}$.
We note that the ratio $\rho_{\mathcal{T}'}(T_\mathrm{RH})/\rho_{\mathcal{T}}(T_\mathrm{RH})$ is nothing to do with the $CP$-violating parameters $\varepsilon_{\mathcal{T}^{(\prime)}}$.
In this figure, we choose the initial PBH mass to be $10^7 \, \mathrm{g}$, $5 \times 10^7 \, \mathrm{g}$, and $3 \times 10^8 \, \mathrm{g}$.
The mass of the color-triplet scalar $\mathcal{T}$ is taken to be $m_\mathcal{T} = 10^6 \,\mathrm{GeV}$.
For a lighter initial PBH mass $M_\mathrm{BH}^\mathrm{in} = 10^7 \, \mathrm{g}$ (blue-thick line), both $\mathcal{T}$ and $\mathcal{T}'$ are produced at the initial stage of the PBH evaporation for the range of $m_{\mathcal{T}'}/m_\mathcal{T} \simeq 0.1 \text{--} 10$. 
Hence, the energy densities are almost independent of the triplet masses within the range of $m_{\mathcal{T}'}/m_\mathcal{T} \simeq 0.1 \text{--} 10$, and the ratio of the energy densities is determined by the ratio of the effective degrees of freedom. 
In this study, we find $\rho_{\mathcal{T}'}(T_\mathrm{RH})/\rho_{\mathcal{T}}(T_\mathrm{RH}) \simeq 1$ in this range since both $\mathcal{T}$ and $\mathcal{T}'$ have the same degrees of freedom. 
For the initial PBH mass of $M_\mathrm{BH}^\mathrm{in} = 10^8 \, \mathrm{g}$ (red-dashed line), both $\mathcal{T}$ and $\mathcal{T}'$ are produced at the early stage of the PBH evaporation as $m_{\mathcal{T}'}/m_\mathcal{T} \lesssim 1$ since the Hawking temperature for the PBH mass of $10^8 \, \mathrm{g}$ is already close to their mass. 
Hence, the ratio of the energy densities approaches to a certain finite value of $\mathcal{O}(1)$ for $m_{\mathcal{T}'}/m_\mathcal{T} \lesssim 1$. 
On the other hand, as $m_{\mathcal{T}'}/m_\mathcal{T} \gtrsim 1$, since $\mathcal{T}'$ is not produced until the Hawking temperature exceeds $m_{\mathcal{T}'}$, $\mathcal{T}'$ is delayed to be produced from the PBH evaporation compared to $\mathcal{T}$. 
Thus, the ratio of the energy densities is strongly suppressed in this region.
Finally, for a heavier mass $M_\mathrm{BH}^\mathrm{in} = 10^9 \, \mathrm{g}$ (green-dotted line), the mass difference between $\mathcal{T}$ and $\mathcal{T}'$ significantly changes the ratio of the energy densities because heavier particles are produced at much later stage of evaporation, and hence the energy density for the heavier particle is strongly suppressed. 

\begin{figure}[t]
  \centering
    \includegraphics[width=0.32\linewidth]{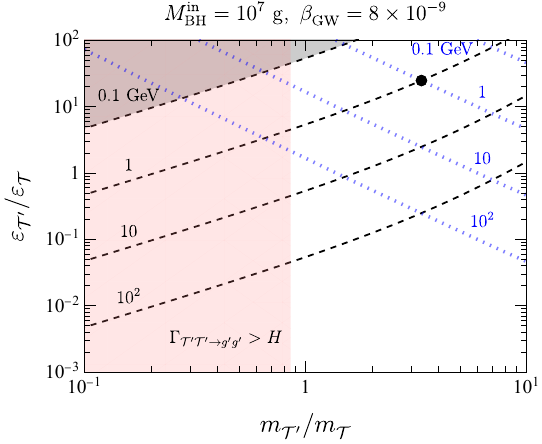}
    \includegraphics[width=0.32\linewidth]{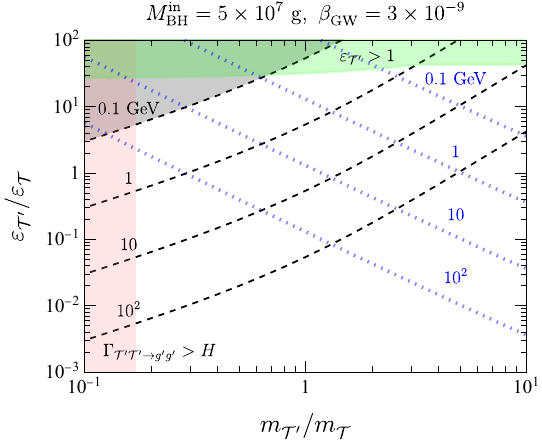}
    \includegraphics[width=0.32\linewidth]{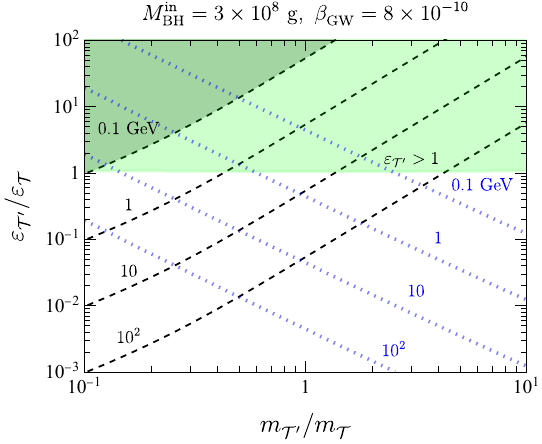}
    \includegraphics[width=0.32\linewidth]{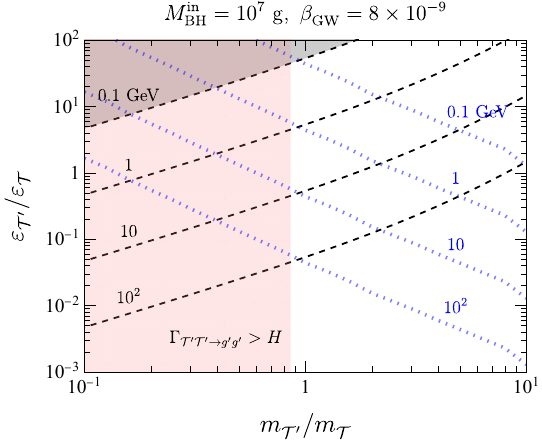}
    \includegraphics[width=0.32\linewidth]{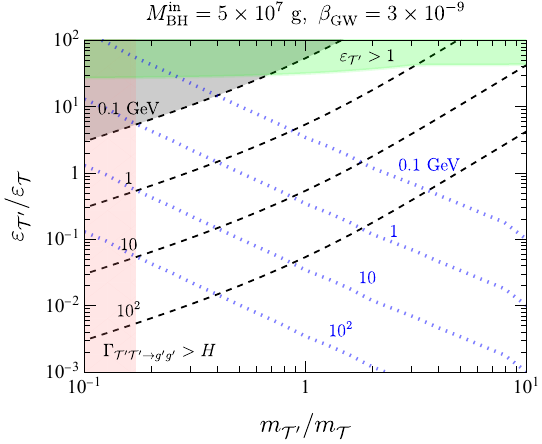}
    \includegraphics[width=0.32\linewidth]{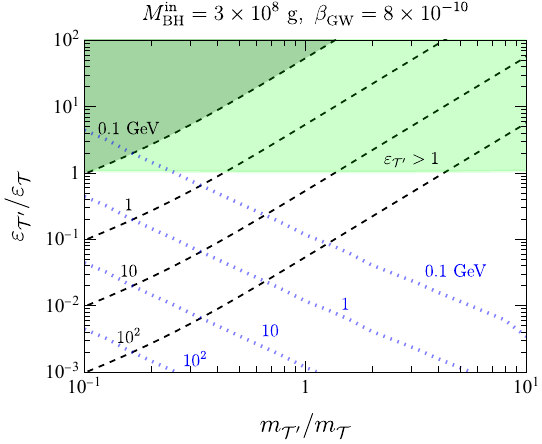}
  \caption{
    The DM mass prediction for the fixed parameters $(m_\mathcal{T}\,,M_\mathrm{BH}^\mathrm{in}\,,\varepsilon_\mathcal{T}\,,\beta)$ from $0.1 \,\mathrm{GeV}$ to $100 \,\mathrm{GeV}$, indicated by black dashed lines: the color-triplet mass of $m_\mathcal{T} = 10^6 \, \mathrm{GeV}$ is fixed and the initial PBH mass of $10^7 \, \mathrm{g}$ (left panels), $5\times 10^7\,\mathrm{g}$ (center panels), and $3\times 10^8 \, \mathrm{g}$ (right panels).
    $\varepsilon_\mathcal{T}$ is fixed to produce the correct relic abundance, $\Omega_B h^2 = 0.022$, for each $m_\mathcal{T}$ and $M_\mathrm{BH}^\mathrm{in}$.
    $\beta$ is taken at the upper bound required by the energy density of gravitational waves at the time of BBN, denoted by $\beta_\mathrm{GW}$.
    The values on each plot show the DM mass in GeV. 
    The blue dotted lines indicate, for each DM mass, the region where the relic abundance of DM produced by the decay of thermally produced dark color-triplets reaches 1\% of the total DM abundance. 
    In the upper panels, the decay width of the dark color-triplet is set to $\Gamma_{\mathcal{T}'}=\Gamma_{\mathcal{T}}$, while in the lower panels, it is set to $\Gamma_{\mathcal{T}'}=10^2\Gamma_{\mathcal{T}}$.
    The assumption that the dark photon is the lightest particle in the dark sector is no longer valid for DM masses below $100 \, \mathrm{MeV}$.
  }
  \label{fig:mDMcontour}
\end{figure}

In \zcref{fig:mDMcontour}, we show the predicted DM mass (in GeV) as a function of the ratios $m_{\mathcal{T}'}/m_{\mathcal{T}}$ and $\varepsilon_{\mathcal{T}'}/\varepsilon_{\mathcal{T}}$, as indicated by the black dashed lines.
We choose the parameters for the baryon asymmetry, $m_{\mathcal{T}} \,,
\varepsilon_{\mathcal{T}}$, to achieve the observed energy density $\Omega_B h^2 = 0.022$ as shown in \zcref{fig:mPBHvsmT}: the parameters $m_{\mathcal{T}}$ and $
\varepsilon_{\mathcal{T}}$ are close to the parameters on the benchmark symbols in \zcref{fig:mPBHvsmT}.
To show the thermal contribution to the DM abundance in the figure, we also choose a specific value for $\beta$: 
$\beta$ is taken to be at the upper bound required by the energy density of gravitational waves at the time of BBN \cite{Domenech:2020ssp},
\begin{align}
\beta_\mathrm{GW}^{} = 1.1\times 10^{-6} \left( \frac{M_\mathrm{BH}^\mathrm{in}}{10^4\,\mathrm{g}} \right)^{-17/24} \,.
\end{align}
The $CP$-violating parameter satisfies $\varepsilon_{\mathcal{T}'} > 1$ in the green-shaded area, while $\mathcal{T}'$ is quickly depleted after the evaporation in the pink-shaded area. 
For the DM mass below $0.1 \, \mathrm{GeV}$ (grey-shaded area in \zcref{fig:mDMcontour}), the assumption that the dark photon is the lightest particle in the dark sector is no longer valid. 
The DM mass is proportional to $\varepsilon_{\mathcal{T}}/\varepsilon_{\mathcal{T}'}$ as shown in \zcref{eq:DMmass}, which can be found in \zcref{fig:mDMcontour}.
Meanwhile, the DM mass is proportional not only to $m_{\mathcal{T}'}/m_{\mathcal{T}}$ but also to $\rho_{\mathcal{T}}/\rho_{\mathcal{T}'}$.
In particular, the ratio of the energy densities changes significantly when the initial PBH mass is heavy enough such that the heavy scalars are not produced in the early stage of evaporation. 
For the heavy PBHs as in the right panels, the DM mass changes steeply following the significant suppression of the energy density. 
The blue dotted line represents, for each DM mass, the region where the contribution from DM produced from the thermal bath reaches 1\% of the total DM abundance (see \zcref{app:thermal_contribution} for details: We use the black dot in the upper-left panel as a benchmark for \zcref{fig:z_vs_Y_rho_S}).
The only difference between the upper and lower panels is the position of the blue line, which reflects the different sizes of the decay width of the dark color-triplet, $\Gamma_{\mathcal{T}'}$. 
In the upper panels, we take $\Gamma_{\mathcal{T}'}=\Gamma_{\mathcal{T}}$, while in the lower panels, we take $\Gamma_{\mathcal{T}'}=10^2\Gamma_{\mathcal{T}}$. 
The region above and to the right of the blue dotted line, along the black dashed line corresponding to the same DM mass, is excluded due to DM overabundance. 
When $\beta$ is smaller than $\beta_{\mathrm{GW}}$, the dilution due to entropy injection becomes weaker.
Consequently, the blue dotted contour shifts toward the lower-left region. 
Meanwhile, the non-thermal contributions shown as the black-dashed lines do not change by the change of $\beta$ nor the change of the decay width $\Gamma_{\mathcal{T}'}$.

We comment on the dark photon mass dependence of the grey-shaded area.
The dark pion mass is naively expected to be an order of magnitude smaller than the dark baryon mass at least. 
Thus, one may think that we can take the dark photon mass an order of magnitude smaller than the DM mass to avoid the overabundance of dark pions.
For the DM mass of $100\,\mathrm{MeV}$, it requires the introduction of the dark photon lighter than about $\mathcal{O}(10)\,\mathrm{MeV}$. 
Meanwhile, the dark photon lighter than $10\,\mathrm{MeV}$ heats up only electron-photon plasma by its decay and annihilation after the neutrino decoupling, and hence it is constrained by the effective number of neutrino degrees of freedom~\cite{Ibe:2019gpv}. 
Therefore, the excluded region is qualitatively the same as the grey-shaded area even if the dark photon mass varies in accordance with the DM mass.

%%%%%%%%%%%%%%%%%%%%%%%%%%%%%%%%%%%%%%%%%%%%%%%%%%%%%%%%%%%%%%%%
\section{Discussion and Conclusion \label{sec:Conc}}

The evaporation of PBHs has the potential to generate particles with masses far above the electroweak scale. 
We have proposed a cogenesis scenario for the composite ADM, where heavy scalar particles produced via PBH evaporation generate the baryon and DM asymmetries in this study. 
The heavy scalar triplet $\mathcal{T}$ which couples to quarks and leptons is constrained to lie within the mass range of $10^6$--$10^9 \, \mathrm{GeV}$ for successful baryogenesis.
The mass of heavy dark scalar $\mathcal{T}'$ must also be within a comparable mass range for successful DM--anti-DM asymmetry generation; a significant mass hierarchy between $\mathcal{T}$ and $\mathcal{T}'$ is disfavored.

In this work, we have assumed a monochromatic PBH mass function. 
A natural extension is to consider an extended mass distribution for the PBH mass.
The initial PBH abundance and the PBH mass are constrained in order to explain the sufficient particle--anti-particle asymmetries for baryon and dark baryon. 
In addition, there may also be the observational bounds on the PBH mass that is irrelevant to the asymmetry production since the PBH mass would be broadly distributed.
Besides, we have not included several effects on the PBH evaporation in this study, such as the memory-burden effect and the formation of hot spots.
Based on quantum-information considerations, the PBH evaporation would be delayed, known as the memory-burden effect, and hence the preferred mass range may differ from that considered in this work.
The PBH evaporation may locally heat up the surrounding plasma, which is called the hot spot \cite{Das:2021wei,He:2022wwy,He:2024wvt}, and the hot spot may affect the asymmetry generation.

It is worthwhile to investigate the UV models for this scenario, since we have a large mass hierarchy between the SM/dark-sector particles and the heavy scalar particles $\mathcal{T}$ and $\mathcal{T}'$. 
One of the candidates is a type of grand unification; however, there remain several problems to address. 
One of them is to suppress the nucleon decay rate as mentioned in the text, and hence the scalars $\mathcal{T}$ must originate from a multiplet different from the SM Higgs doublet, or the Yukawa coupling of $\mathcal{T}$ is quite suppressed by some mechanism compared to the SM Yukawa couplings.
One may consider a scenario with a different heavy scalar particle, which couples to both quark and dark quark: the out-of-equilibrium, $CP$-violating decay of the single scalar particle provides both baryon asymmetry and DM asymmetry. 
Our analysis already covers this scenario: both asymmetries are determined by the heavy scalar parameters $(m_\mathcal{T} \,, \varepsilon_\mathcal{T})$ and the PBH mass as with \zcref{fig:mPBHvsmT}.
Since the number density asymmetries in the two sectors are determined by the same parameter and tightly correlated, the DM mass should be about $5 \, \mathrm{GeV}$ in order to account for the observed ratio of energy densities, $\Omega_\mathrm{DM} h^2/\Omega_B h^2 \simeq 5.45$.

In contrast to models assuming shared asymmetries, we do not need to introduce any portal interactions which convert DM number into baryon number (or lepton number) in the cogenesis scenarios with PBH evaporations. 
Consequently, the DM decay does not occur, and indirect searches via cosmic-ray signals of the DM decay are not applicable to this scenario~\cite{Covi:2009xn,Fukuda:2014xqa,Das:2024tmx}.
Meanwhile, the scenario can be tested through the searches for long-lived particles, where dark baryons and dark mesons can be produced at the terrestrial experiments, such as collider experiments and beam-dump experiments~\cite{Kamada:2021cow,Kuwahara:2023vfc}.

\section*{Acknowledgements}

T.K. thanks Masanori Tanaka for fruitful discussions. 
The work of T.K. is supported in part by the National Science Foundation of China (NSFC) under Grant Nos.~11675002, 11635001, 11725520, 12235001, and the NSFC Research Fund for International Scientists Grant No.~12250410248.
The work of Y.U. is supported in part by NSFC under Grant No.~12347112.

\appendix
\section{Analysis of PBH Evaporation \label{app:anal_PBH}}
In this appendix, we show the analysis of the PBH evaporation, which is used in our numerical computation, in detail.
Once the PBH mass goes below a certain threshold value, the PBH emits all possible particles whose masses are below the Planck scale. 
We can find an analytic solution of the differential equation \zcref{eq:massloss} for $M_\mathrm{BH}$ in this regime since the evaporating function $\epsilon(M_\mathrm{BH})$ is nearly constant.
The PBH mass changes drastically at the final stage of evaporation, leading to the instability of the numerical solver of \zcref{eq:massloss}.
Therefore, we can make the numerical computation stable by matching the numerical solution to the analytic solution.
We match the numerical solution $M_\mathrm{BH} (t)$ to the analytical one denoted by $M_\mathrm{BH}^\mathrm{an}(t)$ at a matching time $t_\mathrm{m}$.
The matching time should be chosen so that the BH temperature at the time is quite larger than the mass of the heaviest particles in the effective models. 
We find the analytic solution as
\begin{align}
    M_\mathrm{BH}(t)
    =
    M_\mathrm{BH}^\mathrm{an}(t)
    \equiv
    3^{1/3}(-c_1 t+c_2)^{1/3} 
    \quad
    (t_\mathrm{m}\leq t \leq t_\mathrm{ev}) \, ,
\end{align}
where $c_1 = 5.34\times 10^{25}\epsilon(0)\,\mathrm{sec}^{-1}$ and $c_2$ is determined by matching the analytic solution with the numerical value at $t_\mathrm{m}$.
We define $t_\mathrm{ev}$ as the time when the PBH mass $M_\mathrm{BH}(t)$ is equal to the Planck mass $4\pi M_\mathrm{Pl}$, in other words, the evaporation is assumed to terminate at this point.
Since we are agnostic about quantum gravity, we cannot deal with the evaporation of PBHs whose masses below the Planck scale. 
We also assume that there is no new particles in addition to the particles introduced in \zcref{sec:PBHsAsym}.
The presence of additional particles would slightly change the temperature evolution of $M_\mathrm{BH}$.

Similarly, the comoving energy density of PBHs $\varrho_\mathrm{BH}$ can be given by the analytical solution of the differential equation \zcref{eq:diffeq-varrhoPBH}. 
In particular, the comoving energy density of PBHs for $t \geq t_\mathrm{m}$ can be written in terms of the analytic solution for the PBH mass $M_\mathrm{BH}^\mathrm{an}$ as follows. 
\begin{align}
    \varrho_\mathrm{BH}(t)
    =
    \varrho^\mathrm{an}_\mathrm{BH}(t)
    \equiv
    \frac{M_\mathrm{BH}^\mathrm{an}(t)}{M_\mathrm{BH}^\mathrm{an}(t_\mathrm{m})}
    \varrho_\mathrm{BH}(t_\mathrm{m})
    \quad
    (t_\mathrm{m}\leq t \leq t_\mathrm{ev}) \, .
\end{align}

We comment on the temperature dependence of the effective degrees of freedom of relativistic particles in the presence of new particles, which is used for the evolution of temperature \eqref{eq:temp_evol}.
We incorporate the contribution of new particles to $g_{*\rho}$ and $g_{*S}$ as the step function as follows.
\begin{align}
    \label{eq:g*rho-BSM}
    g_{*\rho,S}(T)
    =
    g_{*\rho,S}^\mathrm{SM}(T)
    +\sum_{i=\mathrm{bosons}} g_i\theta(T-m_i)
    +\frac{7}{8}\sum_{i=\mathrm{fermions}} g_i\theta(T-m_i) \, ,
\end{align}
where $g_{*\rho,S}^\mathrm{SM}$ denotes the contributions of the SM particles to $g_{*\rho,S}$, and the second term and the third term correspond to the contributions from new particles.
Here, we assume all new particles are in thermal equilibrium with the SM particles. 
In our case, due to the dark confinement, the relativistic degrees of freedom change from dark quarks to dark hadrons as the temperature decreases.
We do not take into account the change of $g_{*\rho,S}$ due to dark confinement, while the similar change due to the QCD confinement is incorporated in $g_{*\rho,S}^\mathrm{SM}$.

\section{On Thermal Contribution \label{app:thermal_contribution}}
We have considered the asymmetry generation only by the heavy scalar particles $\mathcal{T}$ and $\mathcal{T}'$ produced from the PBH evaporation in the text, and we have ignored the thermal contribution of $\mathcal{T}$ and $\mathcal{T}'$ to the asymmetries in the early Universe.
We discuss that the thermal contributions of $\mathcal{T}$ and $\mathcal{T}'$ are not significant in this appendix.

First, the heavy color-triplet scalar $\mathcal{T}$ can be produced in the SM thermal bath during the thermal history since the temperature at the PBH formation is quite high. 
Its CP-violating decay would provide the baryon asymmetry, while $B-L$ is conserved. 
Hence, as far as the initial $B-L$ is zero, this asymmetry is washed out by the sphaleron process until its decoupling. 
The net baryon asymmetry generated by the thermal production of $\mathcal{T}$ is therefore negligible.

Meanwhile, the thermal contribution of the heavy dark scalar $\mathcal{T}'$ to the DM asymmetry cannot be washed out since there is no counterpart of the sphaleron process in the dark sector.
Below the decoupling temperature of the annihilation process and the inverse decay process, the DM asymmetry generated by the thermal production of $\mathcal{T}'$ is frozen.
The DM asymmetry generated by the thermal production of $\mathcal{T}'$ is estimated by solving the coupled Boltzmann equations for the number density of $\mathcal{T}'$ and the DM asymmetry:
\begin{align}
    \frac{d Y_{\mathrm{DM}}}{dx}
    & = \frac{\varepsilon_{\mathcal{T}'} \langle \Gamma_{\mathcal{T}'} \rangle}{x H} \left( Y_{\mathcal{T}'} - Y_{\mathcal{T}'}^\mathrm{eq} \right) \,, \\
    \frac{d Y_{\mathcal{T}'}}{dx}
    & = - \frac{\langle \Gamma_{\mathcal{T}'} \rangle}{x H} \left( Y_{\mathcal{T}'} - Y_{\mathcal{T}'}^\mathrm{eq} \right) - \frac{s \langle \sigma v \rangle}{x H} \left[ Y_{\mathcal{T}'}^2 - (Y_{\mathcal{T}'}^\mathrm{eq})^2 \right] \,.
\end{align}
where $x = m_{\mathcal{T}'}/T$, $Y_{\mathrm{DM}} = (n_{\mathrm{DM}} - n_{\overline{\mathrm{DM}}})/s$, and $Y_{\mathcal{T}'} = n_{\mathcal{T}'}/s$.
$\langle \Gamma_{\mathcal{T}'} \rangle$ denotes the thermally-averaged decay rate of $\mathcal{T}'$, and $\langle \sigma v \rangle$ denotes the thermally-averaged annihilation cross section of $\mathcal{T}'$.
We may have the scattering processes changing the DM number mediated by the heavy scalar $\mathcal{T}'$, while we ignore them for simplicity. 

\begin{figure}[!t]
  \centering
    \includegraphics[width=0.75\linewidth]{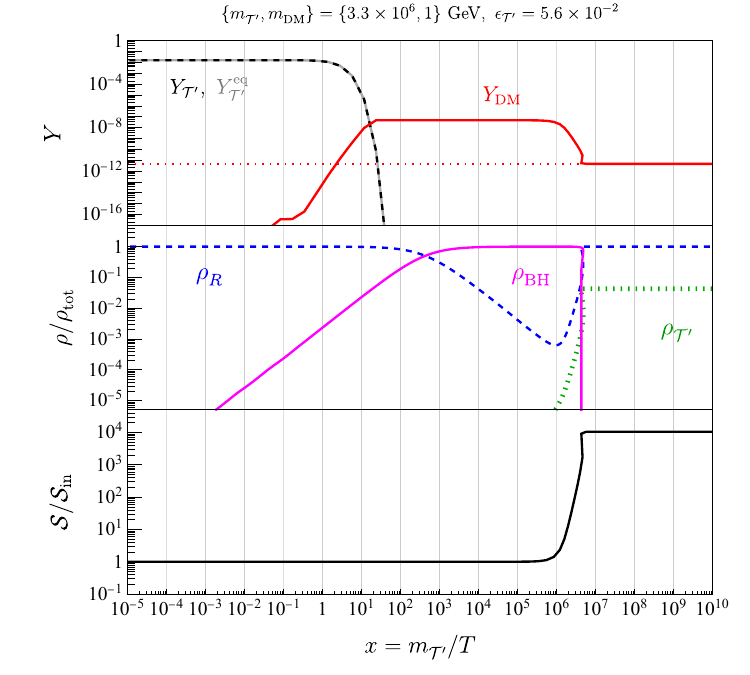}
  \caption{
    Time evolution of the relevant quantities at the benchmark point corresponding to the black dot in the upper-left panel of \zcref{fig:mDMcontour}.
    (\textit{Top}): The black dashed line shows the yield of dark color-triplets produced from the thermal bath, while the gray solid line shows the yield assuming thermal equilibrium. 
    The red solid line represents the DM yield produced through the decay of dark color-triplets generated from the thermal plasma. 
    The red dotted line corresponds to the DM yield for which the relic abundance amounts to 1\% of the total DM abundance.
    (\textit{Middle}): Time evolution of the various energy densities. 
    The blue dashed, magenta solid, and green dotted lines show the energy densities of radiation, PBHs, and DM produced through PBH evaporation, respectively.  
    (\textit{Bottom}): Time evolution of the ratio of the comoving entropy to its initial value.
    }
  \label{fig:z_vs_Y_rho_S}
\end{figure}

\zcref{fig:z_vs_Y_rho_S} shows the time evolution of the relevant quantities at the benchmark point corresponding to the black dot in the upper-left panel of \zcref{fig:mDMcontour}. 
In the top panel, we show the time evolution of the yields of each particle. 
The black dashed line represents the yield of the dark color-triplet $\mathcal{T}'$ produced from the thermal plasma, while the gray solid line shows its yield assuming that $\mathcal{T}'$ remains in thermal equilibrium. 
The red solid line represents the DM yield produced through the decay of the thermally produced $\mathcal{T}'$. 
The red dotted line indicates the DM yield corresponding to 1\% of the total DM abundance.
In the middle panel, the blue dashed, magenta solid, and green dotted lines show the energy densities of radiation, PBHs, and $\mathcal{T}'$ produced through PBH evaporation, respectively. 
In the bottom panel, we show the time evolution of the ratio of the comoving entropy to its initial value.

As shown in the top and the middle panel of \zcref{fig:z_vs_Y_rho_S}, the DM asymmetry is thermally generated before the PBH evaporation.
Meanwhile, the entropy injection can also be significantly effective just after the PBH evaporation as shown in the bottom panel of \zcref{fig:z_vs_Y_rho_S}. 
In this case, the DM asymmetry generated by the thermal production of $\mathcal{T}'$ is diluted by the entropy production from the PBH evaporation.
\zcref{fig:mDMcontour} includes the region where the thermally produced DM asymmetry can be beyond $\mathcal{O}(1)\%$. 
In the remaining region, the DM asymmetry generated by the thermal production of $\mathcal{T}'$ can be safely negligible compared to the DM asymmetry generated by the PBH evaporation.

\section{\texorpdfstring{$\beta$}{beta}-dependence \label{app:beta_dependence}}

\begin{figure}[!t]
  \centering
    \includegraphics[width=0.75\linewidth]{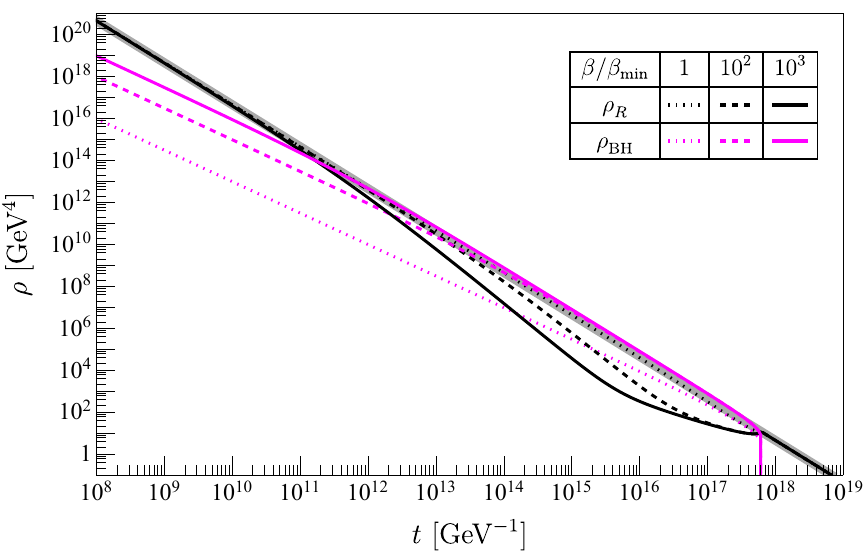}
  \caption{
    Time evolution of $\rho_R$ (black) and $\rho_\mathrm{BH}$ (magenta). 
    Different line types correspond to different $\beta / \beta_\mathrm{min}$. 
    The gray-thick line represents $\rho_R$ in the absence of PBHs, which is proportional to $t^{-2}$.
    We take $M_\mathrm{BH} = 10^7 \, \mathrm{g}$.
    }
  \label{fig:t_vs_rho}
\end{figure}

In this appendix, we discuss how the final result depends on the initial ratio of energy densities $\beta \equiv \rho_\mathrm{BH}(T_0)/\rho_R(T_0)$. 
\zcref{eq:diffeq-varrhoPBH} leads to the simple scaling of the comoving energy density $\varrho_\mathrm{BH} \propto \beta$.
Meanwhile, the dependence of the energy density $\rho_\mathrm{BH}$ on $\beta$ can be found from the equation, 
\begin{align}
    a H \frac{d \rho_\mathrm{BH}}{da} + 3 H \rho_\mathrm{BH} & = \frac{d \ln M_\mathrm{BH}}{dt} \rho_\mathrm{BH} \,.
\end{align}
The loss of the energy density via the evaporation is less effective until the late time of the evaporation, and hence we ignore the impact of the right-hand side in this appendix. 
The second term in the left-hand side is proportional to the Hubble expansion rate, which is proportional to the square root of the energy density: the evolution of $\rho_\mathrm{BH}$ is different in the radiation-dominated epoch and in the matter-dominated epoch. 
The Hubble expansion rate scales as $H \propto a^{-2}$ in the radiation-dominated epoch, while it scales as $H \propto a^{-3/2}$ in the matter-dominated epoch.
The energy density $\rho_\mathrm{BH}$ decreases by the Hubble expansion in the radiation-dominated epoch at the early stage.
Once the PBHs dominate the Universe, $\rho_\mathrm{BH}$ decreases by the Hubble expansion in the matter-dominated epoch.
The Universe quickly enters the matter-dominated epoch when the initial ratio $\beta$ is large. 
$\rho_R$ develops as $\rho_R \simeq \rho_R^0 (t/t_0)^{-2}$ with the initial abundance $\rho_R^0$ and the initial time $t_0$ in the absence of PBHs (the gray-thick line in \zcref{fig:t_vs_rho}). 
In the matter-dominated epoch, the scale factor scales as $a \propto t^{2/3}$, and thus the Hubble rate is given by $H = 2/3t$. 
The energy density of PBHs follows the scaling $\rho_\mathrm{BH} \simeq \rho_\mathrm{BH}^0 (t/t_\mathrm{eq})^{-2}$, where $t_\mathrm{eq}$ denotes the time when the PBH energy density and the radiation energy density are equal, $\rho_R = \rho_\mathrm{BH}$.
We can find that $\rho_\mathrm{BH}$ decreases along the gray-thick line in \zcref{fig:t_vs_rho} in the matter-dominated epoch, as it shares the same time dependence as the radiation energy density in the absence of PBHs.
Therefore, for fixed initial mass $M_\mathrm{BH}^\mathrm{in}$, the PBH energy density does not depend on $\beta$ just before the completion of the PBH evaporation as far as there exists the matter-dominated epoch.

\begin{figure}[!t]
  \centering
    \includegraphics[width=0.75\linewidth]{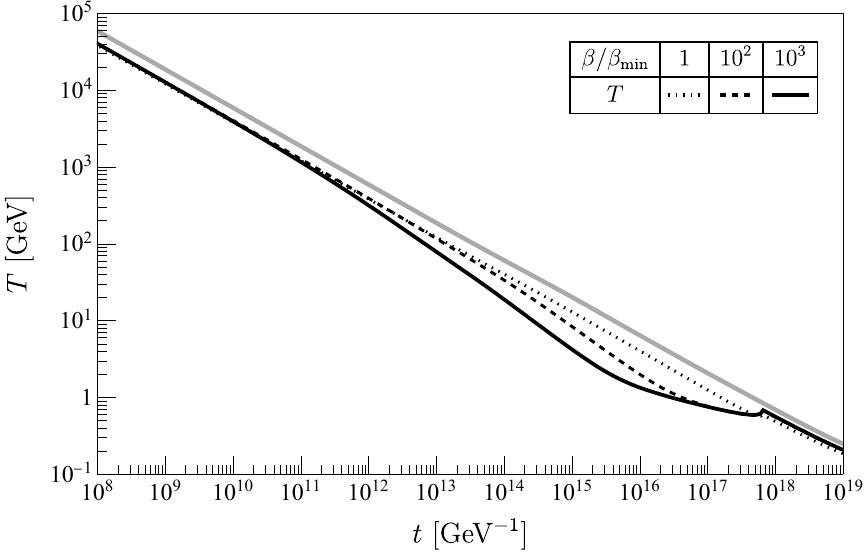}
  \caption{
    Time evolution of $T$. 
    The gray-thick line represents $T$ in the absence of PBHs.
    All other conventions are the same as in \zcref{fig:t_vs_rho}.
    }
  \label{fig:t_vs_T}
\end{figure}

In addition, the entropy density $s$ just before the completion of the PBH evaporation does not strongly depend on $\beta$.
Once we write \zcref{eq:entropy_injection} in terms of the entropy density $s$, the right-hand side is proportional to the energy density of PBHs per the temperature of the thermal plasma $\rho_\mathrm{BH}/T$. 
As stated above, the energy density of PBHs just before the completion of the PBH evaporation does not depend on $\beta$.
The temperature of the thermal plasma just before the completion of the PBH evaporation is also less dependent on $\beta$ as can be seen from \zcref{fig:t_vs_T}: since the Universe is matter-dominated by PBHs at this stage, the reheating temperature is primarily determined by the PBH energy density just before the completion of the PBH evaporation, which is almost independent of $\beta$.
Therefore, the final yields of $\mathcal{T}$ and $\mathcal{T}'$ are less sensitive to the initial ratio of energy densities $\beta$.

Since the Hubble expansion rate depends on $\beta^{1/2}$, the scale factor will also change depending on $\beta$ at the early stage of the matter-dominated epoch.
The comoving energy density $\varrho_\mathrm{BH}$ is proportional to $\beta$ as mentioned above, while the energy density $\rho_\mathrm{BH}$ is independent from $\beta$ in the end.
Therefore, the scale factor follows a non-trivial scaling at the late stage of the matter-dominated epoch, $a \propto \beta^{1/3}$.

\bibliographystyle{utphys}
\bibliography{ref}
\end{document}